\documentclass[twocolumn]{revtex4}
\usepackage{mathbbold}
\usepackage{amsfonts}
\usepackage{amsmath}
\usepackage{amssymb}
\usepackage{charter}
\usepackage{graphicx}

\begin{document}

\title{Multi-photon-addition amplified coherent state}
\author{Xue-feng Zhan$^{1}$, Qiang Ke$^{1}$, Min-xiang Li$^{2}$, and
Xue-xiang Xu$^{1,\dag }$ }
\affiliation{$^{1}$College of Physics and Communication Electronics, Jiangxi Normal
University, Nanchang 330022, China;\\
$^{2}$School of Education, Jiangxi Normal University, Nanchang 330022, China;%
\\
$^{\dag }$xuxuexiang@jxnu.edu.cn}

\begin{abstract}
State $g^{\hat{n}}\hat{a}^{\dag m}\left\vert \alpha \right\rangle $ and
state $\hat{a}^{\dag m}g^{\hat{n}}\left\vert \alpha \right\rangle $ are same
to state $\hat{a}^{\dag m}\left\vert g\alpha \right\rangle $, which is
called as multi-photon-addition amplified coherent state (MPAACS) by us.
Here, $\hat{n}$, $\hat{a}^{\dag }$, $\left\vert \alpha \right\rangle $, $g$ (%
$\geq 1$), and $m$ are photon number operator, creation operator, coherent
state, gain facor, and an interger, respectively. We study mathematical and
physical properties for these MPAACSs, including normalization, photon
component analysis, Wigner function, effective gain, quadrature squeezing,
and equivalent input noise. Actually, the MPAACS, which contains more
nonclassicality, is an amplified version of photon-added coherent state
(PACS) introduced by Agrwal and Tara [Phys. Rev. A 43, 492 (1991)]. Our work
provides theoretical references for implementing amplifiers for light fields.

\textbf{Keywords: }noiseless linear amplification; repeated photon addition;
amplified coherent state; noise; Wigner function
\end{abstract}

\maketitle

\section{Introduction}

In physics, signal amplification is a simple concept. However, signal
amplification unavoidably comes with noise\cite{1}. Often, the introduced
noise makes it difficult for people to distinguish the amplified signal. The
downside is that quantum noise will restrict quantum technologies such as
quantum cloning\cite{2,3} and superluminal information transfer\cite{4}. In
order to conquer this restriction, people are more looking forward to taking
advantage of noiseless amplification, which can be implemented by
probabilistic operations\cite{5,6}. In recent years, the theoretical and
experimental study on ideal noiseless linear amplification (NLA) has
attracted the interests of the researchers\cite{7,8,9,10}. The NLA has been
used for many quantum information tasks such as loss suppression\cite{11},
quantm repeater\cite{12}, quantum error correction\cite{13}, and
entanglement distillation\cite{14}.

The NLA can be described by the operator $g^{\hat{n}}$ (AM), where $\hat{n}=%
\hat{a}^{\dag }\hat{a}$ is the photon number operator and $g$ ($g>1$) is the
gain factor. Here $g^{\hat{n}}\equiv \hat{1}$ is just the identity operator
if $g=1$. The NLA\ can be implemented probabilistically by\ combinating
multiple photon addition and subtraction with current technology\cite{15}.
Theoretically, an ideal noiseless amplifier (described by $g^{\hat{n}}$) can
map an input state $\rho _{in}$ into an output state $\rho _{out}$, i.e., $%
g^{\hat{n}}:\rho _{in}\longmapsto \rho _{out}$. For input Fock state $%
\left\vert n\right\rangle $, we have $g^{\hat{n}}:\left\vert n\right\rangle
\longmapsto \left\vert n\right\rangle $ due to $\hat{n}\left\vert
n\right\rangle =n\left\vert n\right\rangle $. For input coherent state (CS),
we have $g^{\hat{n}}:\left\vert \alpha \right\rangle \longmapsto \left\vert
g\alpha \right\rangle $ because of $g^{\hat{n}}\left\vert \alpha
\right\rangle =e^{(g^{2}-1)\left\vert \alpha \right\vert ^{2}/2}\left\vert
g\alpha \right\rangle $. Generally, the input states to be amplified by $g^{%
\hat{n}}$ include Gaussian and non-Gaussian state\cite{16}. Moreover, the
output states must be re-normalized after operating $g^{\hat{n}}$ because
this operator\ is unbounded.

In 2016, Park et al. suggested and compared several schemes for
non-deterministic NLA of CSs using $\hat{a}^{\dag 2}$, $\hat{a}\hat{a}^{\dag
}$, $\left( \hat{a}\hat{a}^{\dag }\right) ^{2}$, $\hat{a}^{\dag 4}$, $\hat{a}%
\hat{a}^{\dag }\hat{a}^{\dag 2}$, $\hat{a}^{\dag 2}\hat{a}\hat{a}^{\dag }$,
which may work as amplifiers for CSs with weak, medium or large amplitudes.
Among them, the two-photon addition ($\hat{a}^{\dag 2}$) scheme work more
effectively than others as a noiseless amplifier\cite{17}. Before then, in
1991, Argarwal and Tara had introduced photon-added CSs (PACS) $\hat{a}%
^{\dag m}\left\vert \alpha \right\rangle $, which can be generated by
applying a $m$-photon addition $\hat{a}^{\dag m}$ (AD)\ on the CS $%
\left\vert \alpha \right\rangle $, where $m$ is an interger\cite{18}. It is
undeniable that the AD $\hat{a}^{\dag m}$ also works as an amplifier to
amplify $\left\vert \alpha \right\rangle $\ to $\hat{a}^{\dag m}\left\vert
\alpha \right\rangle $.

In this paper, we shall study the behaviour of $\hat{a}^{\dag m}\left\vert
\alpha \right\rangle $\ under the action of $g^{\hat{n}}$. In another words,
we shall obtain new quantum states by further applying $g^{\hat{n}}$ on\ $%
\hat{a}^{\dag m}\left\vert \alpha \right\rangle $. We will analyze their
nonclassical properties and amplification effects. The key is to examine\
the combinatorial effects of AD $\hat{a}^{\dag m}$\ and AM $g^{\hat{n}}$ on
CSs. The remaining paper is organized as follows. In Sec.2, we introduce our
considered states accompanying with description and normalization. In Sec.3,
we analyze photon components for them. In Sec.4, we study Wigner functions
to show character of Gaussianity and nonclassicality. Sec.5 is devoted to
studying effective gain, quadrature squeezing and equivalent input noise for
these states. Conclusions are summarized in the last section.

\section{Generating quantum states}

Based on operators $g^{\hat{n}}$, $\hat{a}^{\dag m}$, and CS $\left\vert
\alpha \right\rangle $, we introduce a class of new states by two equivalent
ways. The conceptual generating schemes are shown in Fig.1.
\begin{figure}[tbp]
\label{Fig1} \centering\includegraphics[width=0.95\columnwidth]{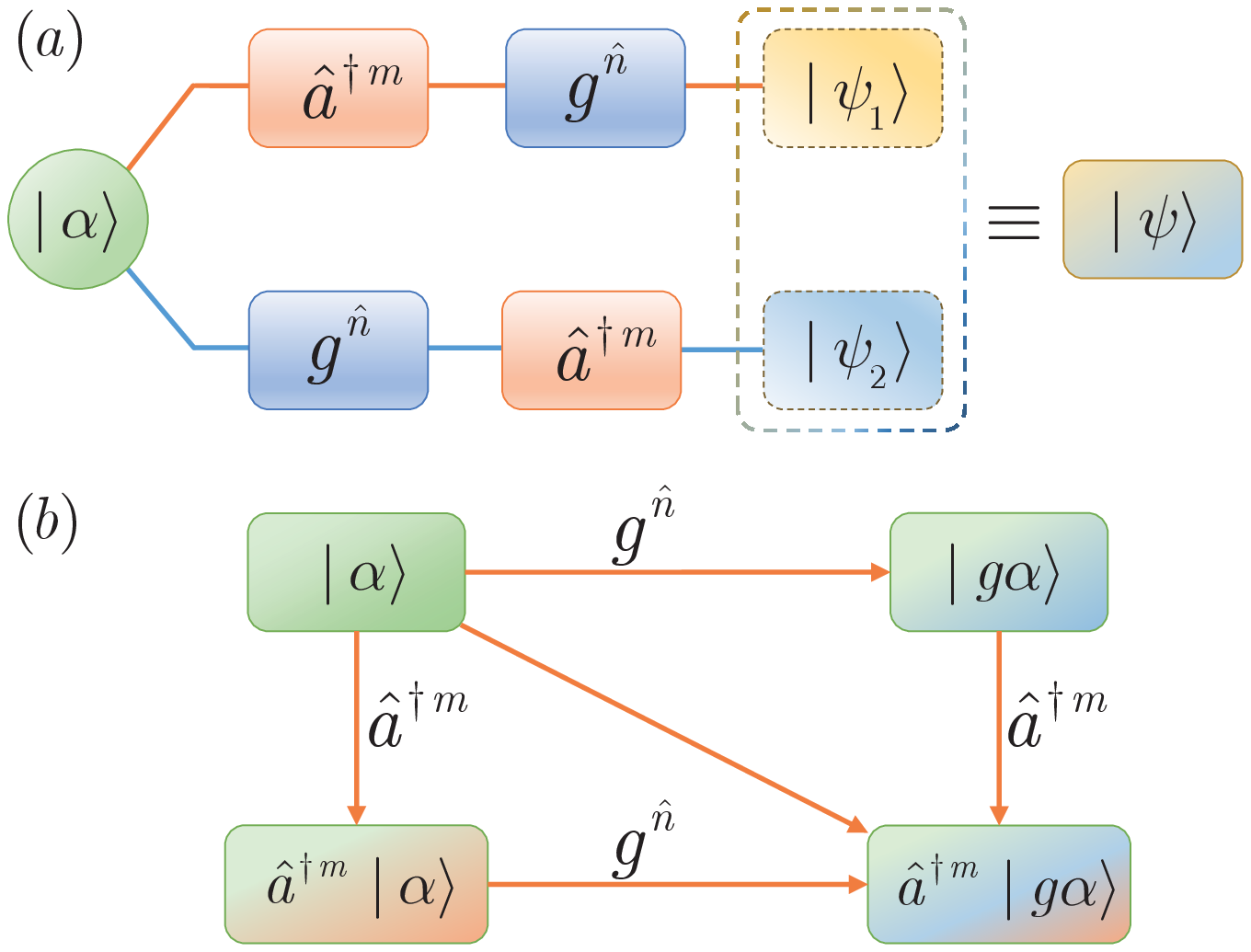}
\caption{(a) Conceptional generating schemes of ADAMCS $\left\vert \protect%
\psi _{1}\right\rangle $ and AMADCS $\left\vert \protect\psi %
_{2}\right\rangle $. It is interesting to prove that $\left\vert \protect%
\psi _{1}\right\rangle $\ and $\left\vert \protect\psi _{2}\right\rangle $
are the same MPAACS $\left\vert \protect\psi \right\rangle $. (b) The map $%
\left\vert \protect\alpha \right\rangle \longmapsto a^{\dag m}\left\vert g%
\protect\alpha \right\rangle $ can be implemented by way 1 $\left\vert
\protect\alpha \right\rangle \longmapsto a^{\dag m}\left\vert \protect\alpha %
\right\rangle \longmapsto a^{\dag m}\left\vert g\protect\alpha \right\rangle
$ or by way 2 $\left\vert \protect\alpha \right\rangle \longmapsto
\left\vert g\protect\alpha \right\rangle \longmapsto a^{\dag m}\left\vert g%
\protect\alpha \right\rangle $ after employing $g^{\hat{n}}$ and $\hat{a}%
^{\dag m}$ in sequence.}
\end{figure}

\textit{Way 1:} Employing $\hat{a}^{\dag m}$ then $g^{\hat{n}}$ on $%
\left\vert \alpha \right\rangle $, we get the addition-amplification
coherent state (ADAMCS)
\begin{equation}
\left\vert \psi _{1}\right\rangle =\frac{1}{\sqrt{N_{1}}}g^{\hat{n}}\hat{a}%
^{\dag m}\left\vert \alpha \right\rangle ,  \label{1-1}
\end{equation}%
with normalization factor
\begin{equation}
N_{1}=g^{2m}m!e^{\left( g^{2}-1\right) \left\vert \alpha \right\vert
^{2}}L_{m}(-g^{2}\left\vert \alpha \right\vert ^{2}),  \label{1-2}
\end{equation}%
where $L_{m}(x)$\ is the $m$th-order Laguerre polynomial\cite{19}. This map
include two processes, i.e., $\hat{a}^{\dag m}:$ $\left\vert \alpha
\right\rangle \longmapsto \hat{a}^{\dag m}\left\vert \alpha \right\rangle $
and $g^{\hat{n}}:$ $\hat{a}^{\dag m}\left\vert \alpha \right\rangle
\longmapsto \left\vert \psi _{1}\right\rangle $, where $\hat{a}^{\dag
m}\left\vert \alpha \right\rangle $\ is the intermediate state.

\textit{Way 2:} Employing $g^{\hat{n}}$ then $\hat{a}^{\dag m}$ on $%
\left\vert \alpha \right\rangle $, we get the amplification-addition
coherent state (AMADCS)%
\begin{equation}
\left\vert \psi _{2}\right\rangle =\frac{1}{\sqrt{N_{2}}}\hat{a}^{\dag m}g^{%
\hat{n}}\left\vert \alpha \right\rangle ,  \label{1-3}
\end{equation}%
with normalization factor
\begin{equation}
N_{2}=m!e^{\left( g^{2}-1\right) \left\vert \alpha \right\vert
^{2}}L_{m}(-g^{2}\left\vert \alpha \right\vert ^{2}).  \label{1-4}
\end{equation}%
This map include two processes, i.e., $g^{\hat{n}}:$ $\left\vert \alpha
\right\rangle \longmapsto \left\vert g\alpha \right\rangle $ and $\hat{a}%
^{\dag m}:$ $\left\vert g\alpha \right\rangle \longmapsto \left\vert \psi
_{2}\right\rangle $, where $\left\vert g\alpha \right\rangle $\ is the
intermediate state.

Although operators $g^{\hat{n}}$ and $\hat{a}^{\dag m}$ do not commute with
each other (i.e. $g^{\hat{n}}\hat{a}^{\dag m}\neq \hat{a}^{\dag m}g^{\hat{n}}
$), the relation $g^{\hat{n}}\hat{a}^{\dag m}=g^{m}\hat{a}^{\dag m}g^{\hat{n}%
}$ is satisfied and leads to $\left\vert \psi _{1}\right\rangle =\left\vert
\psi _{2}\right\rangle $ after normalization. Thus, we redefine them as $%
\left\vert \psi \right\rangle $ in the form%
\begin{equation}
\left\vert \psi \right\rangle =\frac{1}{\sqrt{N}}\hat{a}^{\dag m}\left\vert
g\alpha \right\rangle ,  \label{1-5}
\end{equation}%
with normalization factor $N=m!L_{m}(-g^{2}\left\vert \alpha \right\vert
^{2})$. In this paper, we call $\left\vert \psi \right\rangle $\ as MPAACS,
which is just an amplified PACS. By using $\left\vert \psi _{1}\right\rangle
$\ in Appendix A or by using $\left\vert \psi _{2}\right\rangle $\ in
Appendix B in two parallel ways, we get the expressions of state
description, normalization facor, expectation value, density matrix
elements, and Wigner function for $\left\vert \psi \right\rangle $. Without
question, the results in Appendix A are same to those in Appendix B except $%
N_{1}=g^{2m}N_{2}$. As illustrated in Table I, states including $\left\vert
0\right\rangle $, $\left\vert m\right\rangle $, $\left\vert \alpha
\right\rangle $, $\left\vert g\alpha \right\rangle $, and $\hat{a}^{\dag
m}\left\vert \alpha \right\rangle $ are special cases of $\left\vert \psi
\right\rangle $ with proper $g$, $\alpha $, and $m$.
\begin{table}[h]
\caption{Special cases of the MPAACS}
\begin{center}
\begin{tabular}{|c|c|c|}
\hline\hline
$\left\vert \psi \right\rangle $ & parameter conditions & state name \\
\hline\hline
$\left\vert 0\right\rangle $ & $m=0,\alpha =0$ & vacuum state \\ \hline
$\left\vert m\right\rangle $ & $\alpha =0$ & Fock state \\ \hline
$\left\vert \alpha \right\rangle $ & $g=1,m=0$ & coherent state \\ \hline
$\left\vert g\alpha \right\rangle $ & $m=0$ & amplified CS \\ \hline
$a^{\dag m}\left\vert \alpha \right\rangle $ & $g=1$ & photon-added CS \\
\hline
\end{tabular}%
\end{center}
\end{table}

\section{Component analysis of the MPAACS}

The MPAACS can be expanded\ in terms of Fock states as%
\begin{equation}
\left\vert \psi \right\rangle =\sum_{k=m}^{\infty }c_{k}\left\vert
k\right\rangle ,  \label{2-1}
\end{equation}%
with%
\begin{equation}
c_{k}=\frac{\sqrt{k!}\left( g\alpha \right) ^{k-m}e^{-g^{2}\left\vert \alpha
\right\vert ^{2}/2}}{\left( k-m\right) !\sqrt{m!L_{m}(-g^{2}\left\vert
\alpha \right\vert ^{2})}}.  \label{2-2}
\end{equation}%
It is interesting to note that the components including $\left\vert
k\right\rangle $ ($k<m$) are missing in the MPAACS. Moreover, when $g=1$,
the result can be reduced to that in the work of Agarwal and Tara\cite{18}.
Accordingly, the density matrix element (DME) for $\rho =\left\vert \psi
\right\rangle \left\langle \psi \right\vert $ can be expressed as $\rho
_{kl}=\left\langle k\right\vert \rho \left\vert l\right\rangle
=c_{k}c_{l}^{\ast }$, whose numerical results also can be calculated
according to Eq.(A.7) or Eq.(B.6) in the Appendix. The photon number
distribution (PND) can be written as $\rho _{kk}=\left\vert c_{k}\right\vert
^{2}$, i.e., the diagonal terms of the density matrix. Without loss of
generality, we shall take $\alpha =\left\vert \alpha \right\vert e^{i\theta
_{p}}$ only with $\theta _{p}=0$ in the numerical analysis.

Fig.2 presents the PNDs $\rho _{kk}$ of the MPAACSs with different
parameters ($\left\vert \alpha \right\vert $, $g$, $m$). The results show
that: (a) The effect of photon-added number $m$ can be observed in Fig.2(a)
at fixed $\left\vert \alpha \right\vert =1$, $g=2$ and for three different $%
m $. As $m$ increases, the PNDs $\rho _{kk}$ approach higher-photon regime
where all the $\left\vert k\right\rangle $\ terms with $k<m$\ are missing,
like a shifted version of original amplified CS ($m=0$). (b) The effect of
gain factor $g$ can be observed in Fig.2(b) at fixed $\left\vert \alpha
\right\vert =1$, $m=2$ and for three different $g$. It is obvious to see
that the PNDs approach higher-photon regime as $g$\ increasing. (c) The
effect of the field amplitude $\left\vert \alpha \right\vert $ can be
observed in Fig.2(c) at fixed $g=2$, $m=2$ and for three different $%
\left\vert \alpha \right\vert $. We find if $\left\vert \alpha \right\vert
=0 $, there is the sole component of $\left\vert m\right\rangle $. If case $%
\left\vert \alpha \right\vert >0$, the PNDs approach higher-photon regime as
$\left\vert \alpha \right\vert $\ increasing.

Fig.3 presents the absolute values of the DMEs ($\left\vert \rho
_{kl}\right\vert $) of the MPAACSs showing the effects of parameters ($%
\left\vert \alpha \right\vert $, $g$, $m$). Generally, two adjacent states,
corresponding to two adjacent graphs in Fig.3, can be connected through a
map with its operation. These maps include $\hat{a}^{\dag m}:\left\vert
0\right\rangle \longmapsto \left\vert m\right\rangle $ for (a)-(b); $D\left(
\alpha \right) :\left\vert 0\right\rangle \longmapsto \left\vert \alpha
\right\rangle $ for (a)-(c), where $D\left( \alpha \right) =e^{\alpha
a^{\dag }-\alpha ^{\ast }a}$ is the displacement operator; $\hat{a}^{\dag
m}:\left\vert \alpha \right\rangle \longmapsto \hat{a}^{\dag m}\left\vert
\alpha \right\rangle $ for (c)-(d); $g^{\hat{n}}:\left\vert \alpha
\right\rangle \longmapsto \left\vert g\alpha \right\rangle $ for (c)-(e); $%
g^{\hat{n}}:\hat{a}^{\dag m}\left\vert \alpha \right\rangle \longmapsto \hat{%
a}^{\dag m}\left\vert g\alpha \right\rangle $ for (d)-(f); $\hat{a}^{\dag
m}:\left\vert g\alpha \right\rangle \longmapsto \hat{a}^{\dag m}\left\vert
g\alpha \right\rangle $ for (e)-(f). Of course, the map $\left\vert
m\right\rangle \longmapsto \hat{a}^{\dag m}\left\vert \alpha \right\rangle $
for (b)-(d) can not be realized simply by $D\left( \alpha \right) $. It can
be easily seen from Fig.3 that: (1) The AD $\hat{a}^{\dag m}$ leads to the
rescale of corrresponding DMEs and the displacement to higher indeces ($\rho
_{k,l}\rightarrow \rho _{k+m,l+m}$), leaving all $\rho _{k,l}$ with $k,l<m$
void. (2) The AM $g^{\hat{n}}$ leads to the rescale of corrresponding DMEs.
\
\begin{figure*}[tbp]
\label{Fig2} \centering\includegraphics[width=1.8\columnwidth]{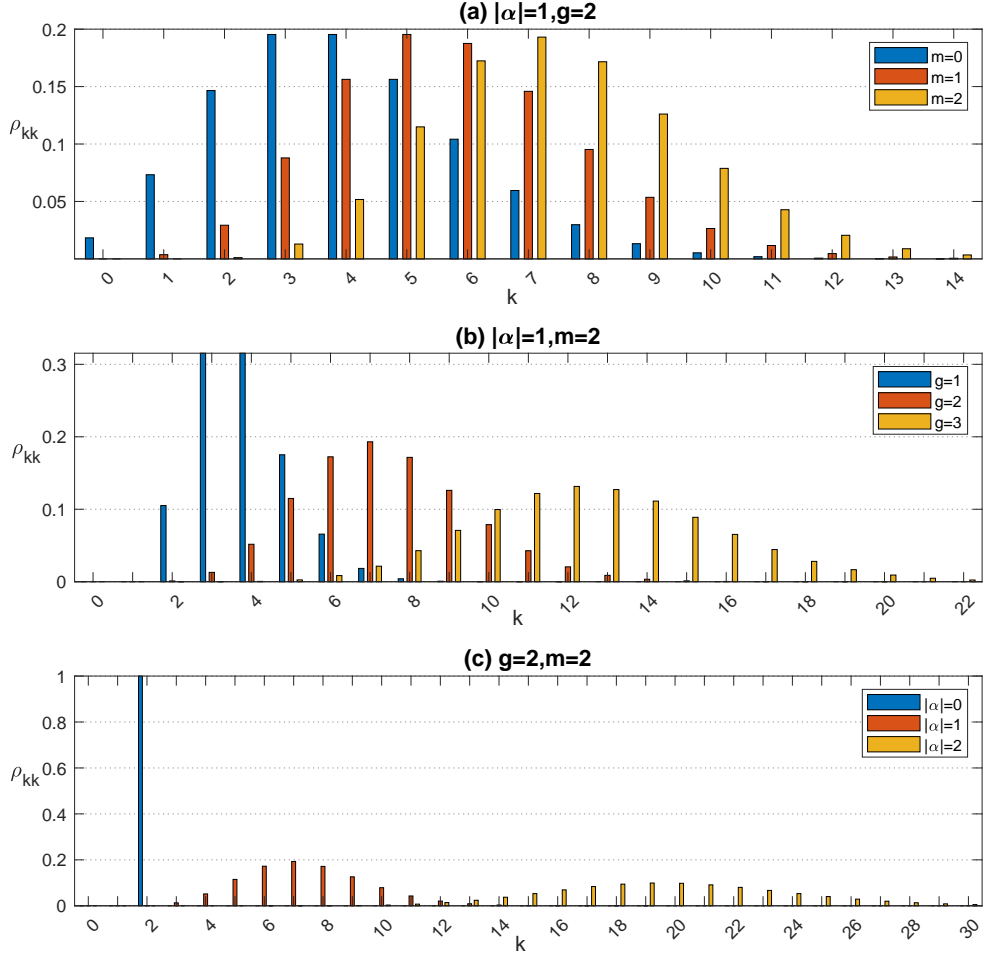}
\caption{PNDs of $\left\vert \protect\psi \right\rangle $ with different ($%
\left\vert \protect\alpha \right\vert $, $g$, $m$). (a) showing the effect
of $m$; (b) showing the effect of $g$; (c) showing the effect of $\left\vert
\protect\alpha \right\vert $.}
\end{figure*}
\begin{figure*}[tbp]
\label{Fig3} \centering\includegraphics[width=1.8\columnwidth]{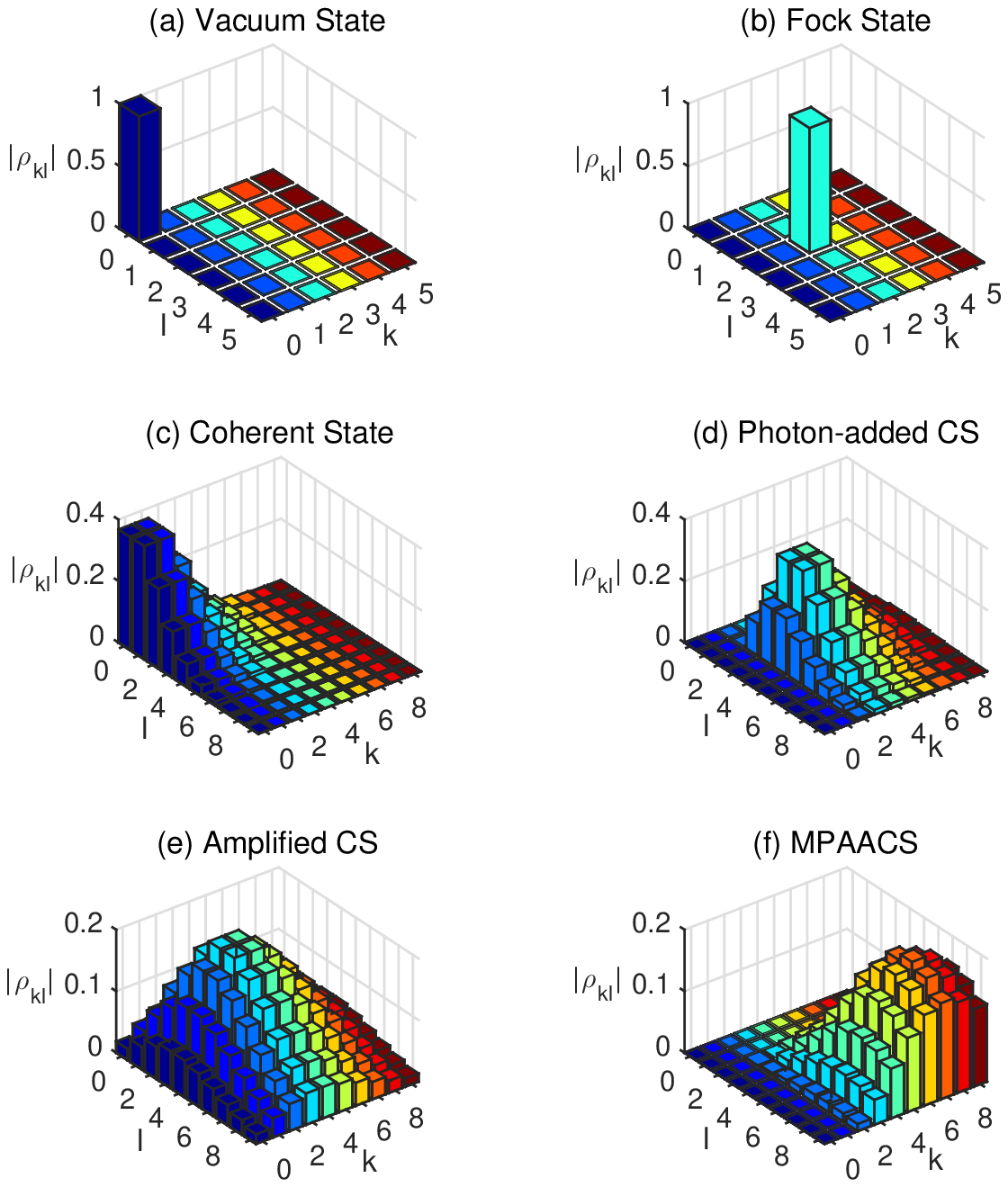}
\caption{Absolute values of DMEs for $\left\vert \protect\psi \right\rangle $%
-($\left\vert \protect\alpha \right\vert $, $g$, $m$) taking (a) $\left\vert
0\right\rangle $-($0$, $1$, $0$); (b) $\left\vert m\right\rangle $-($0$, $1$%
, $2$); (c) $\left\vert \protect\alpha \right\rangle $-($1$, $1$, $0$); (d) $%
a^{\dag m}\left\vert \protect\alpha \right\rangle $-($1$, $1$, $2$); (e) $%
\left\vert g\protect\alpha \right\rangle $-($1$, $2$, $0$); and (f) $a^{\dag
m}\left\vert g\protect\alpha \right\rangle $-($1$, $2$, $2$). Indeed, any
two adjacent states can be connected through appropriate operations such as $%
D\left( \protect\alpha \right) $, $a^{\dag m}$, and $g^{\hat{n}}$.}
\end{figure*}

\section{Wigner function of the MPAACS}

In phase-space formalism, the Wigner function (WF) is an important
quasiprobability distribution, representing the corresponding quantum state%
\cite{20,21,22}. One can judge Gaussianity or non-Gaussianity from its WF
form and non-classicality from its Wigner negativity\cite{23,24}.
Theoretically, $W_{\rho }\left( \beta \right) $ can be also obtained by
means of the following transformation%
\begin{equation}
W_{\rho }\left( \beta \right) =\sum_{k,l=0}^{\infty }\rho _{kl}W_{\left\vert
k\right\rangle \left\langle l\right\vert }\left( \beta \right)  \label{4-0}
\end{equation}%
which is associated with $\rho _{kl}$ and $W_{\left\vert k\right\rangle
\left\langle l\right\vert }\left( \beta \right) $ (WF of operator $%
\left\vert k\right\rangle \left\langle l\right\vert $). But it is very
difficult to simulate perfectly as given by Eq.(\ref{4-0}) because of the
infinite summation. In experiment, the approximate WF can be reconstructed
from a truncated density matrix with finite dimension through tomographic
analysis.

Fortunately, analytical WF of the MPAACS can be obtained from Eq.(A.8) or
Eq.(B.7) as follows%
\begin{equation}
W_{\rho }\left( \beta \right) =\dfrac{2(-1)^{m}L_{m}(\left\vert 2\beta
-g\alpha \right\vert ^{2})}{\pi L_{m}(-g^{2}\left\vert \alpha \right\vert
^{2})}e^{-2\left\vert \beta -g\alpha \right\vert ^{2}},  \label{4-1}
\end{equation}%
in the ($x,y$) phase space, where $\beta =(x+iy)/\sqrt{2}$. As expected, Eq.(%
\ref{4-1}) can reduce to the following special cases.

(1) If $\alpha =0$\ and $m=0$, then we obtain WF of vacuum state
\begin{equation}
W_{\left\vert 0\right\rangle }\left( \beta \right) =\dfrac{2}{\pi }%
e^{-2\left\vert \beta \right\vert ^{2}}.  \label{4-2}
\end{equation}

(2) If $m=0$ and $g=1$, then we obtain WF of coherent state
\begin{equation}
W_{\left\vert \alpha \right\rangle }\left( \beta \right) =\dfrac{2}{\pi }%
e^{-2\left\vert \beta -\alpha \right\vert ^{2}}.  \label{4-3}
\end{equation}

(3) If $m=0$, then we obtain WF of amplified CS
\begin{equation}
W_{\left\vert g\alpha \right\rangle }\left( \beta \right) =\dfrac{2}{\pi }%
e^{-2\left\vert \beta -g\alpha \right\vert ^{2}}.  \label{4-4}
\end{equation}

(4) If $\alpha =0$, then we obtain WF of Fock state
\begin{equation}
W_{\left\vert m\right\rangle }\left( \beta \right) =\dfrac{2}{\pi }%
(-1)^{m}e^{-2\left\vert \beta \right\vert ^{2}}L_{m}(4\left\vert \beta
\right\vert ^{2}).  \label{4-5}
\end{equation}

(5) If $g=1$, then we obtain WF of photon-added CS
\begin{equation}
W_{\hat{a}^{\dag m}\left\vert \alpha \right\rangle }\left( \beta \right) =%
\dfrac{2(-1)^{m}L_{m}(\left\vert 2\beta -\alpha \right\vert ^{2})}{\pi
L_{m}(-\left\vert \alpha \right\vert ^{2})}e^{-2\left\vert \beta -\alpha
\right\vert ^{2}}.  \label{4-6}
\end{equation}%
Eq.(\ref{4-6}) is just the equation (3.8) in Ref.\cite{18}.

Corresponding to Fig.3 and according to Eqs.(\ref{4-1})-(\ref{4-6}), we plot
WFs in Fig.4 and their sections (with $y=0$) in Fig.5(a), as well as their
marginal distributions in Fig.5(b). Here, the marginal distribution\ in $x$\
direction\ can be evaluated numerically by%
\begin{equation}
p\left( x\right) =\int_{-\infty }^{\infty }W\left( \beta \right) dy,
\label{4-7}
\end{equation}%
where the scaling relations such as $\int_{-\infty }^{\infty }W\left( \beta
\right) d^{2}\beta =1$\ and $\int_{-\infty }^{\infty }p\left( x\right) dx=1$
must be ensured.
\begin{figure*}[tbp]
\label{Fig4} \centering\includegraphics[width=2.0\columnwidth]{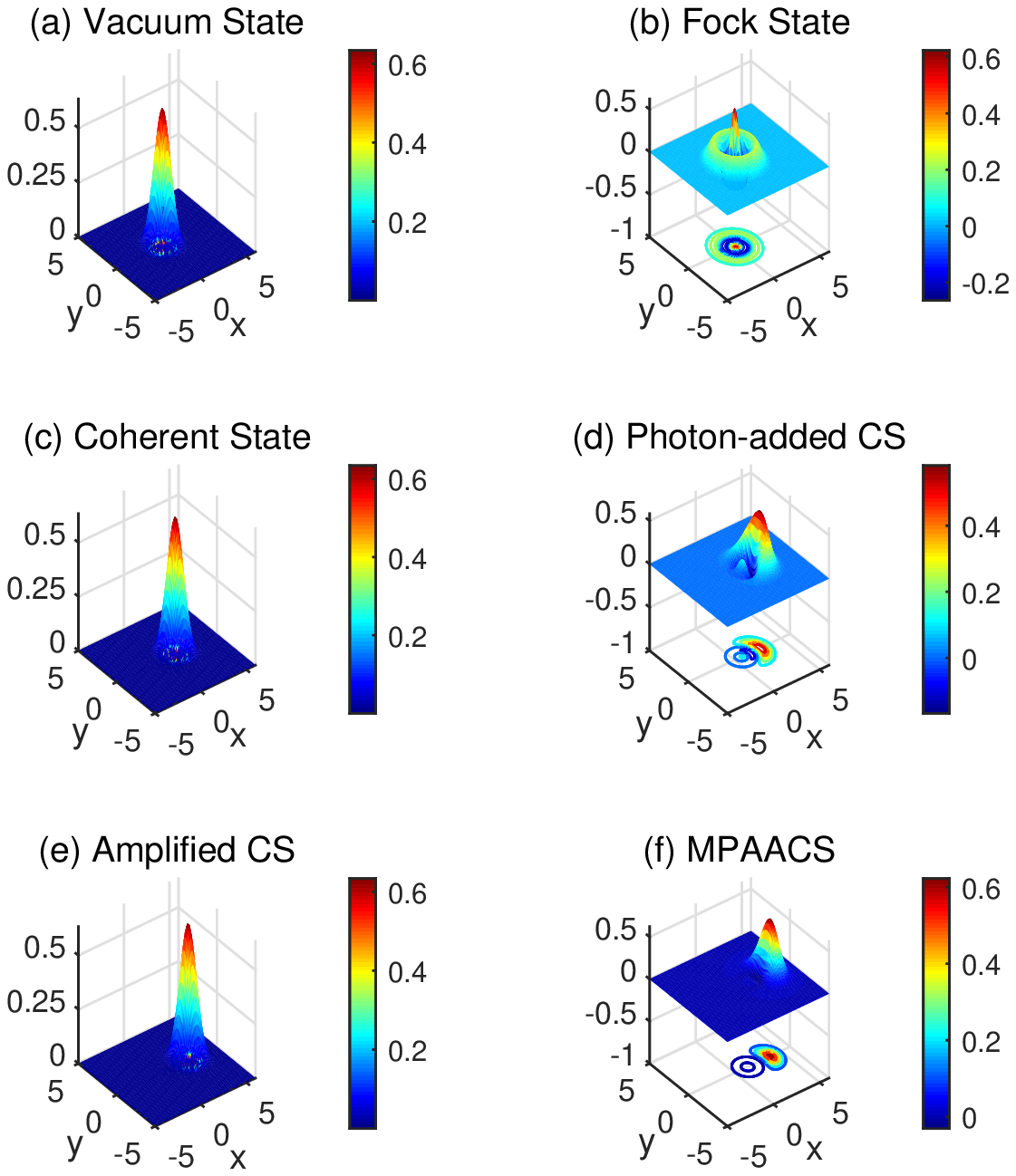}
\caption{WFs $W\left( \protect\beta \right) $ of $\left\vert \protect\psi %
\right\rangle $ in the phase space $(x,y)$. Here, $\left\vert \protect\psi %
\right\rangle $-($\left\vert \protect\alpha \right\vert $, $g$, $m$) are
taking (a) $\left\vert 0\right\rangle $-($0$, $1$, $0$); (b) $\left\vert
m\right\rangle $-($0$, $1$, $2$); (c) $\left\vert \protect\alpha %
\right\rangle $-($1$, $1$, $0$); (d) $a^{\dag m}\left\vert \protect\alpha %
\right\rangle $-($1$, $1$, $2$); (e) $\left\vert g\protect\alpha %
\right\rangle $-($1$, $2$, $0$); and (f) $a^{\dag m}\left\vert g\protect%
\alpha \right\rangle $-($1$, $2$, $2$).}
\end{figure*}
\begin{figure}[tbp]
\label{Fig5} \centering\includegraphics[width=1.0\columnwidth]{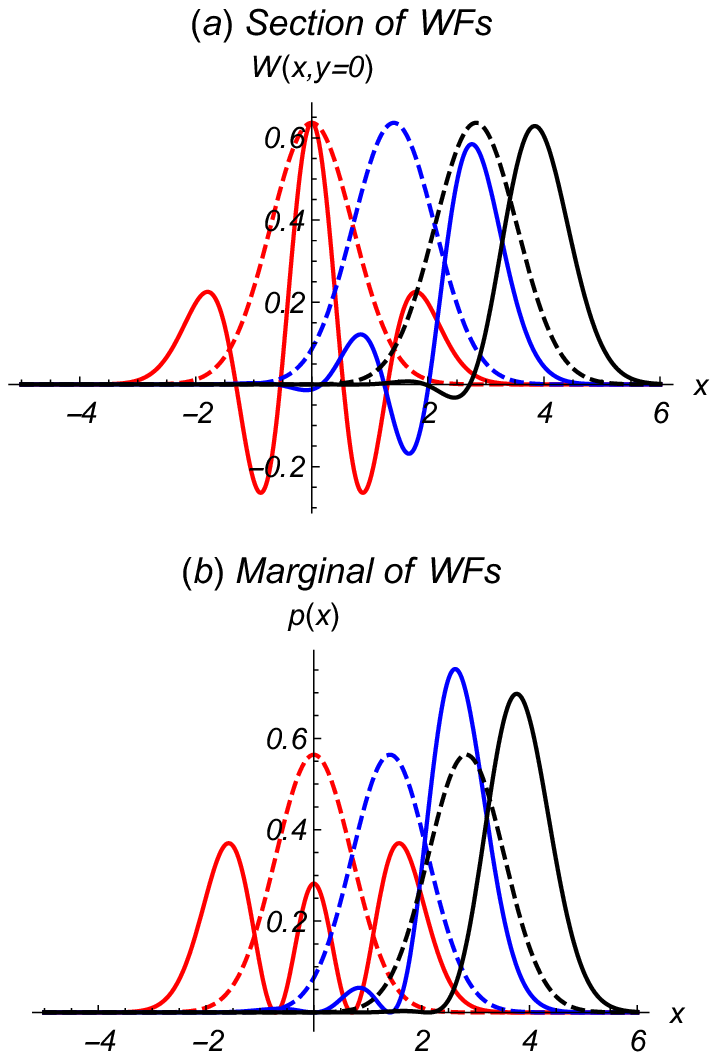}
\caption{(a) Sections $W(x,y=0)$ and (b) Marginal distribution $p(x)$ of $%
W\left( \protect\beta \right) $ for $\left\vert \protect\psi \right\rangle $%
\ in Fig.4 as a function of $x$. Here, $\left\vert \protect\psi %
\right\rangle $-($\left\vert \protect\alpha \right\vert $, $g$, $m$) taking
(red dashed) $\left\vert 0\right\rangle $-($0$, $1$, $0$); (red solid) $%
\left\vert m\right\rangle $-($0$, $1$, $2$); (blue dashed) $\left\vert
\protect\alpha \right\rangle $-($1$, $1$, $0$); (blue solid) $a^{\dag
m}\left\vert \protect\alpha \right\rangle $-($1$, $1$, $2$); (black dashed) $%
\left\vert g\protect\alpha \right\rangle $-($1$, $2$, $0$); and (black
solid) $a^{\dag m}\left\vert g\protect\alpha \right\rangle $-($1$, $2$, $2$%
). }
\end{figure}

The left three figures of Fig.4 correspond to vacuum state $\left\vert
0\right\rangle $, coherent state $\left\vert \alpha \right\rangle $, and
amplified CS $\left\vert g\alpha \right\rangle $. Their distributions are
Gaussian without Wigner negativity and with different central positions. The
right three figures of Fig.4 correspond to Fock state $\left\vert
m\right\rangle $, Photon-added CS $\hat{a}^{\dag m}\left\vert \alpha
\right\rangle $, and MPAACS $\hat{a}^{\dag m}\left\vert g\alpha
\right\rangle $. Their distributions are non-Gaussian and showing Wigner
negativity. Main characters including Gaussianity (or non-Gaussianity) and
nonclassicality (negativity) can also be seen from their corresponding
sections in $y=0$ in Fig.5(a). Morover, we see from Fig.5(b) that, as $m$\
is increased, the width of distribution $p(x)$ becomes narrower comparing to
that of $\left\vert \alpha \right\rangle $\ and $\left\vert g\alpha
\right\rangle $ (corresponding to $m=0$) with $\left\vert \alpha \right\vert
>0$. We think, this is owe to the effect of AD $\hat{a}^{\dag m}$.

\section{Effective gain, quadrature squeezing and equivalent input noise}

Generally, many properties of light states are related to amplitude
quadrature $\hat{x}=\left( a+a^{\dag }\right) /\sqrt{2}$ and phase
quadrature $\hat{p}=\left( a-a^{\dag }\right) /(i\sqrt{2})$\cite{25,26}. In
order to study the combinatorial contributions of AM $g^{\hat{n}}$ and AD $%
\hat{a}^{\dag m}$, we compare the properties of $\left\vert \alpha
\right\rangle $\ and $\left\vert \psi \right\rangle $\ in terms of the
effective gain, quadrature squeezing, and equivalent input noise. After our
full consideration, we only use quadrature operator $\hat{x}$ to discuss all
these properties.

\textit{Effective gain:} An effective gain\cite{27,28} from the input $%
\left\vert \alpha \right\rangle $ to the output $\left\vert \psi
\right\rangle $ can be defined as the ratio of the expectation values of the
quadrature operator $\hat{x}$:
\begin{equation}
g_{eff}=\frac{\left\langle \hat{x}\right\rangle _{\left\vert \psi
\right\rangle }}{\left\langle \hat{x}\right\rangle _{\left\vert \alpha
\right\rangle }}.  \label{3-1}
\end{equation}%
In particularly, we have
\begin{eqnarray}
g_{eff}^{(m=0)} &=&g,  \notag \\
g_{eff}^{(m=1)} &=&g\frac{2+g^{2}\left\vert \alpha \right\vert ^{2}}{%
1+g^{2}\left\vert \alpha \right\vert ^{2}},  \notag \\
g_{eff}^{(m=2)} &=&g\frac{6+6g^{2}\left\vert \alpha \right\vert
^{2}+g^{4}\left\vert \alpha \right\vert ^{4}}{2+4g^{2}\left\vert \alpha
\right\vert ^{2}+g^{4}\left\vert \alpha \right\vert ^{4}},  \notag \\
&&\vdots  \label{3-2}
\end{eqnarray}%
\begin{figure}[tbp]
\label{Fig6} \centering\includegraphics[width=1.0\columnwidth]{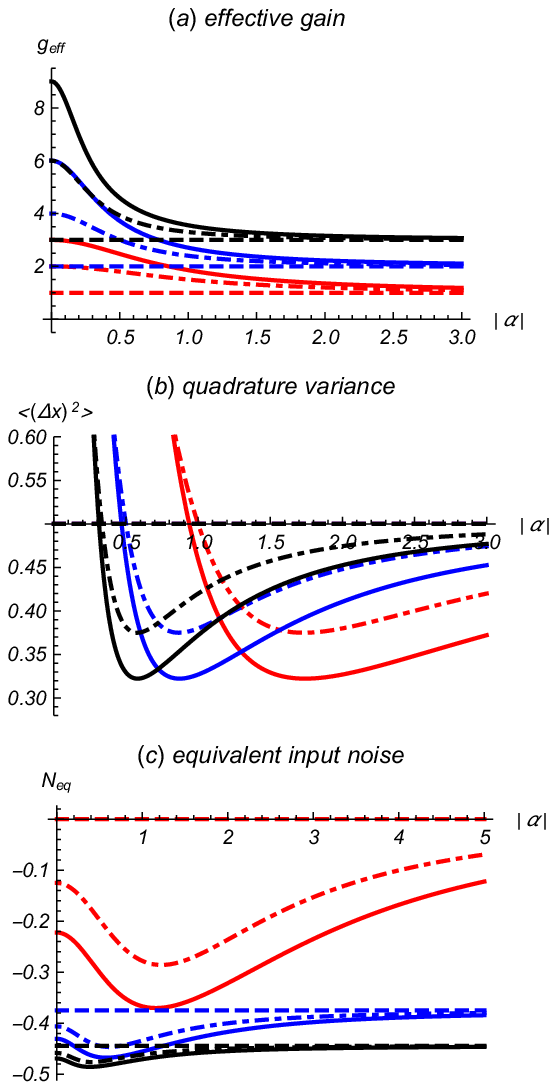}
\caption{(a) Effective gain $g_{eff}$, (b) Variances $\left\langle \left(
\Delta x\right) ^{2}\right\rangle $, and (c) Equivalent input noise $N_{eq}$
as the function of $\left\vert \protect\alpha \right\vert $ . Here,
parameters are taking $g=1$ (red)$,2$ (blue)$,3$ (black) and $m=0$ (dashed)$%
,1$ (dotdashed)$,2$ (solid).}
\end{figure}
In Fig.6(a), we plot $g_{eff}$\ as a function of $\left\vert \alpha
\right\vert $ by taking different $g$\ and $m$. From Fig.6(a), we find that $%
g_{eff}\gtrsim g$\ is always right and $g_{eff}$\ is a monotonical
decreasing function of $\left\vert \alpha \right\vert $ for different $g$
and $m>0$. Two limiting cases include: (1) $g_{eff}\rightarrow (m+1)g$ in
the limit of $\left\vert \alpha \right\vert \rightarrow 0$; (2) $%
g_{eff}\rightarrow g$ in the limit of $\left\vert \alpha \right\vert
\rightarrow \infty $.

\textit{Quadrature squeezing:} Quadrature squeezing, as a nonclassical
character of light states, can be evident by measuring the quadrature
variances\cite{29,30}. The quadrature variance in $\hat{x}$ can be
calculated from
\begin{equation}
\left\langle \left( \Delta \hat{x}\right) ^{2}\right\rangle =\left\langle
\hat{x}^{2}\right\rangle -\left\langle \hat{x}\right\rangle ^{2},
\label{3-3}
\end{equation}%
which is also called as the quadrature flunctuation. Similar definition $%
\left\langle \left( \Delta \hat{p}\right) ^{2}\right\rangle $ is available
to quadrature $\hat{p}$. It is well known that $\left\vert \alpha
\right\rangle $\ and $\left\vert g\alpha \right\rangle $ are not quadrature
squeezing states because of $\left\langle \left( \Delta \hat{x}\right)
^{2}\right\rangle =\left\langle \left( \Delta \hat{p}\right)
^{2}\right\rangle =0.5$. While if $\left\langle \left( \Delta \hat{x}\right)
^{2}\right\rangle $\ or $\left\langle \left( \Delta \hat{p}\right)
^{2}\right\rangle $\ is smaller than $0.5$, then the quantum state is
quadrature squeezing. Thus, a question naturally arises: is the MPAACS
quadrature squeezing? In order to answer this question, we only plot $%
\left\langle \left( \Delta \hat{x}\right) ^{2}\right\rangle $ as a function
of $\left\vert \alpha \right\vert $ for different $\left\vert \psi
\right\rangle $\ in Fig.6(b) regardless of $\left\langle \left( \Delta \hat{p%
}\right) ^{2}\right\rangle \geq 0.5$.

Clearly, we find that (1) For $m=0$ case, the $\hat{x}$\ quadrature variance
remains constant fluctuation $0.5$ (without quadrature squeezing) for any $g$%
\ and $\left\vert \alpha \right\vert $. This is because $\left\vert \psi
\right\rangle $ has been reduced to $\left\vert \alpha \right\rangle $\ or $%
\left\vert g\alpha \right\rangle $ in this case. (2) For $m=1$ case, the $%
\hat{x}$\ quadrature exhibits squeezing if $\left\vert g\alpha \right\vert
>1 $, which is consistent with the result for SPACS\cite{31}. (3) For $m=2$
case, the $\hat{x}$\ quadrature exhibits squeezing if $\left\vert g\alpha
\right\vert >0.938744$. Other similar reduced fluctuations will be exhibited
squeezing if $\left\vert g\alpha \right\vert >0.900407$ for $m=3$, $%
\left\vert g\alpha \right\vert >0.873904$ for $m=4$, $\left\vert g\alpha
\right\vert >0.854454$ for $m=5$, and so on. (4) All above results show that
the MPAACSs except $m=0$ will exhibit quadrature squeezing, only when $%
\left\vert g\alpha \right\vert $ exceeds a certain threshold. (5) Moreover,
we always have $\left\langle \left( \Delta \hat{x}\right) ^{2}\right\rangle
\rightarrow 0.5$ in the limit of $\left\vert \alpha \right\vert \rightarrow
\infty $.

\textit{Equivalent input noise:} Noise of the output $\left\vert \psi
\right\rangle $\ referring to the input $\left\vert \alpha \right\rangle $
may be analyzed in terms of the equivalent input noise (EIN):
\begin{equation}
N_{eq}=\frac{\left\langle \left( \Delta \hat{x}\right) ^{2}\right\rangle
_{\left\vert \psi \right\rangle }}{g_{eff}^{2}}-\left\langle \left( \Delta
\hat{x}\right) ^{2}\right\rangle _{\left\vert \alpha \right\rangle },
\label{3-4}
\end{equation}%
which tells how much noise has been added to the input noise level. In fact,
the EIN came from a classical electronics terminology and used to quantify
the performance of an amplifier\cite{27,28,32,33}. For an amplification
process, the EIN is negative and indicates the characteristic of noiseless
amplification\cite{34}.

In Fig.6(c), the EINs\ are shown as a function of $\left\vert \alpha
\right\vert $ for $\left\vert \psi \right\rangle $ with different $g$\ and $%
m $. The numerical results reveal that $N_{eq}$ is clearly negative\ for all
$\left\vert \alpha \right\vert $, except case $m=0$ and $g=1$. The main
results include: (1) For $m=0$, we know that $N_{eq}$ remains constant $%
0.5/g^{2}$ $-0.5$ for any $g$ and all $\left\vert \alpha \right\vert $; (2)
In the limit of $\left\vert \alpha \right\vert \rightarrow 0$, we see $%
N_{eq}\rightarrow 0.5/g^{2}$ $-0.5$ for $m=0$, $0.375/g^{2}$ $-0.5$ for $m=1$%
, $0.277778/g^{2}$ $-0.5$ for $m=2$, respectively; (3) In the limit of $%
\left\vert \alpha \right\vert \rightarrow \infty $, we see $%
N_{eq}\rightarrow $ $0.5/g^{2}$ $-0.5$ for any $g$ ($>1$)\ and different $m$%
; (4) A minimum value of $N_{eq}$\ can be see at proper $\left\vert \alpha
\right\vert $ for each case.

\section{Conclusion and discussion}

We have introduced MPAACSs by applying AM $g^{\hat{n}}$\ and AD $\hat{a}%
^{\dag m}$\ on $\left\vert \alpha \right\rangle $ and proved that state $g^{%
\hat{n}}\hat{a}^{\dag m}\left\vert \alpha \right\rangle $ and state $\hat{a}%
^{\dag m}g^{\hat{n}}\left\vert \alpha \right\rangle $ are state $\hat{a}%
^{\dag m}\left\vert g\alpha \right\rangle $. From $\left\vert \alpha
\right\rangle $ to $\hat{a}^{\dag m}\left\vert g\alpha \right\rangle $, the
combinatorial effect of $g^{\hat{n}}$\ and $\hat{a}^{\dag m}$ is working as
an amplifier. From the point of view of quantum state engineering, these
MPAACSs $\hat{a}^{\dag m}\left\vert g\alpha \right\rangle $ are a class of
new quantum states, which include many familiar quantum states, such as $%
\left\vert 0\right\rangle $, $\left\vert m\right\rangle $, $\left\vert
\alpha \right\rangle $, $\left\vert g\alpha \right\rangle $, and $\hat{a}%
^{\dag m}\left\vert g\alpha \right\rangle $. We have derived the
normalization factor for the MPAACS and found that it is related to Lagurrel
polynomials. Interesting physical properties are given analytically and
simulated numerically according to the supplementary materials. The main
results are summarized as follows.

As for the effects of AM $g^{\hat{n}}$\ and AD $\hat{a}^{\dag m}$ on photon
components of the MPAACSs, we find that: (1) The AD $\hat{a}^{\dag m}$ leads
to the void of the low-photon components (including $\left\vert
0\right\rangle $, $\left\vert 1\right\rangle $, $\cdots $, $\left\vert
m-1\right\rangle $) and the re-layout of photon components. (2) The AM $g^{%
\hat{n}}$\ leads to the re-layout of photon components. As for the effects
of AM $g^{\hat{n}}$\ and AD $\hat{a}^{\dag m}$ on WFs, we find that: (1) The
AD $\hat{a}^{\dag m}$ is a non-Gaussian operation, which can transform a
Gaussian state into a non-gaussian state, accompanying with Wigner
negativity. (2) The AM $g^{\hat{n}}$ is a Gaussian operation, which can
remain original Gaussianity or non-Gaussianity of quantum state. As for the
effects of AM $g^{\hat{n}}$\ and AD $\hat{a}^{\dag m}$ on amplification,
squeezing and noise, we find that: (1) Both $g^{\hat{n}}$\ and $\hat{a}%
^{\dag m}$ can improve the effective gain by changing $g$\ and $m$. (2) The
quadrature squeezing will exhibit when $\left\vert g\alpha \right\vert $
exceeds a certain threshold except $m=0$. (3) The EINs except case $g=1$\
and $m=0$ are negative, showing the characteristic of noiseless
amplification.

In our previous works\cite{35,36,37}, we have introduced several amplified
quantum states, such as amplified coherent state, amplified thermal state,
and amplified squeezed vacuum, by applying $\left( g-1\right) \hat{n}+1$\ or
$(g-\sqrt{2g-1})\hat{n}^{2}+(\sqrt{2g-1}-1)\hat{n}+1$\ on the corresponding
input states. These amplified states can exhibit their respective peculiar
nonclassicality. Of course, these operators work as amplifiers of realizing
signal amplification. However, it is actually impossible to implement a
perfect noiseless amplifier described by $g^{\hat{n}}$, albeit with zero
success probability\cite{38}. So, our present work only provides a
theoretical reference for signal amplification in quantum technology or
state generation in quantum state engineering.

\section*{Appendix: Supplemental Materials}

Using techniques such as $g^{\hat{n}}=:e^{(g-1)\hat{a}^{\dag }\hat{a}}:$ ($%
:\cdots :$ denotes normal ordering) and $x^{n}=\partial _{s}^{n}e^{sx}|_{s=0}
$, we provide the following supplemental materials for all informations
discussed in the main text.

\subsection*{Appendix A: Information for state $\left\vert \protect\psi %
_{1}\right\rangle $}

In this appendix, we provide the state description, normalization,
expectation value, density matrix elements and Wigner function for $%
\left\vert \psi _{1}\right\rangle $.

\textbf{(a) State description}

Eq.(\ref{1-1}) can be further written as
\begin{subequations}
\begin{equation}
\left\vert \psi _{1}\right\rangle =\frac{e^{-\frac{\left\vert \alpha
\right\vert ^{2}}{2}}}{\sqrt{N_{1}}}\partial _{s_{1}}^{m}e^{g\left( \alpha
+s_{1}\right) a^{\dag }}\left\vert 0\right\rangle |_{s_{1}=0},  \tag{A.1}
\end{equation}%
accompanying with conjugate state
\end{subequations}
\begin{subequations}
\begin{equation}
\left\langle \psi _{1}\right\vert =\frac{e^{-\frac{\left\vert \alpha
\right\vert ^{2}}{2}}}{\sqrt{N_{1}}}\partial _{t_{1}}^{m}\left\langle
0\right\vert e^{g\left( \alpha ^{\ast }+t_{1}\right) a}|_{t_{1}=0},
\tag{A.2}
\end{equation}%
which leads to density operator $\rho _{1}=\left\vert \psi _{1}\right\rangle
\left\langle \psi _{1}\right\vert $,
\end{subequations}
\begin{subequations}
\begin{equation}
\rho _{1}=\frac{e^{-\left\vert \alpha \right\vert ^{2}}}{N_{1}}\partial
_{s_{1}}^{m}\partial _{t_{1}}^{m}e^{g\left( \alpha +s_{1}\right) a^{\dag
}}\left\vert 0\right\rangle \left\langle 0\right\vert e^{g\left( \alpha
^{\ast }+t_{1}\right) a}|_{\left( s_{1},t_{1}\right) =0}.  \tag{A.3}
\end{equation}

\textbf{(b) Normalization}

The normalization factor is
\end{subequations}
\begin{subequations}
\begin{equation}
N_{1}=e^{\left( g^{2}-1\right) \left\vert \alpha \right\vert ^{2}}\partial
_{s_{1}}^{m}\partial _{t_{1}}^{m}e^{g^{2}\left( t_{1}\alpha +s_{1}\alpha
^{\ast }+\allowbreak s_{1}t_{1}\right) }|_{\left( s_{1},t_{1}\right) =0}.
\tag{A.4}
\end{equation}%
Using the following formula
\end{subequations}
\begin{equation}
L_{m}\left( xy\right) =\frac{(-1)^{m}}{m!}\partial _{s}^{m}\partial
_{t}^{m}e^{-st+sx+ty}|_{\left( s,t\right) =0},  \tag{A.5}
\end{equation}%
we easily obtain the analytical result of $N_{1}$ in Eq.(\ref{1-2}).

\textbf{(c) Expectation value}

Here, we give $\langle a^{\dagger k}a^{l}\rangle _{\rho _{1}}$\ as follows
\begin{subequations}
\begin{align}
\langle a^{\dagger k}a^{l}\rangle _{\rho _{1}}& =\frac{e^{\left(
g^{2}-1\right) \left\vert \alpha \right\vert ^{2}}}{N_{1}}\partial
_{s_{1}}^{m}\partial _{t_{1}}^{m}\partial _{f_{1}}^{k}\partial _{h_{1}}^{l}
\notag \\
& e^{g\left( h_{1}\alpha +f_{1}\alpha ^{\ast }\right) +g^{2}\left(
t_{1}\alpha +s_{1}\alpha ^{\ast }+\allowbreak s_{1}t_{1}\right) \allowbreak }
\notag \\
& e^{g\left( h_{1}s_{1}+f_{1}t_{1}\right) }|_{\left(
s_{1},t_{1},f_{1},h_{1}\right) =0}.  \tag{A.6}
\end{align}

\textbf{(d) Density matrix elements}

Here, we give $\rho _{kl}^{(1)}=\left\langle k|\rho _{1}|l\right\rangle $
with the following form
\end{subequations}
\begin{subequations}
\begin{align}
\rho _{kl}^{(1)}& =\frac{e^{-\left\vert \alpha \right\vert ^{2}}}{N_{1}\sqrt{%
k!l!}}\partial _{s_{1}}^{m}\partial _{t_{1}}^{m}\partial
_{f_{1}}^{k}\partial _{h_{1}}^{l}  \notag \\
& e^{gf_{1}\left( \alpha +s\right) +gh_{1}\left( \alpha ^{\ast }+t\right)
}|_{\left( s_{1},t_{1},f_{1},h_{1}\right) =0}.  \tag{A.7}
\end{align}

\textbf{(e) Wigner function}

WF of $\rho _{1}$ has the following form
\end{subequations}
\begin{subequations}
\begin{align}
& W_{\rho _{1}}\left( \beta \right)  \notag \\
& =\dfrac{2}{\pi N_{1}}e^{-\left( g^{2}+1\right) |\alpha |^{2}-2\left\vert
\beta \right\vert ^{2}+2g\allowbreak \alpha \beta ^{\ast }+2g\alpha ^{\ast
}\beta }  \notag \\
& \partial _{s_{1}}^{m}\partial _{t_{1}}^{m}e^{-g^{2}\left( t_{1}\alpha
+s_{1}\alpha ^{\ast }+s_{1}t_{1}\right) +\allowbreak 2g\left( t_{1}\beta
+s_{1}\beta ^{\ast }\right) }|_{\left( s_{1},t_{1}\right) =0}  \tag{A.8}
\end{align}

\subsection*{Appendix B: Information for state $\left\vert \protect\psi %
_{2}\right\rangle $}

In this appendix, we provide the state description, normalization,
expectation value, density matrix elements and Wigner function for $%
\left\vert \psi _{2}\right\rangle $.

\textbf{(a) State description}

Eq.(\ref{1-3}) can be further written as
\end{subequations}
\begin{subequations}
\begin{equation}
\left\vert \psi _{2}\right\rangle =\frac{e^{-\frac{\left\vert \alpha
\right\vert ^{2}}{2}}}{\sqrt{N_{2}}}\partial _{s_{2}}^{m}e^{\left(
s_{2}+g\alpha \right) a^{\dag }}\left\vert 0\right\rangle |_{s_{2}=0},
\tag{B.1}
\end{equation}%
accompanying with conjugate state
\end{subequations}
\begin{subequations}
\begin{equation}
\left\langle \psi _{2}\right\vert =\frac{e^{-\frac{\left\vert \alpha
\right\vert ^{2}}{2}}}{\sqrt{N_{2}}}\partial _{t_{2}}^{m}\left\langle
0\right\vert e^{\left( t_{2}+g\alpha ^{\ast }\right) a}|_{t_{2}=0},
\tag{B.2}
\end{equation}%
and leads to density operator $\rho _{2}=\left\vert \psi _{2}\right\rangle
\left\langle \psi _{2}\right\vert $,
\end{subequations}
\begin{subequations}
\begin{equation}
\rho _{2}=\frac{e^{-\left\vert \alpha \right\vert ^{2}}}{N_{2}}\partial
_{s_{2}}^{m}\partial _{t_{2}}^{m}e^{\left( s_{2}+g\alpha \right) a^{\dag
}}\left\vert 0\right\rangle \left\langle 0\right\vert e^{\left(
t_{2}+g\alpha ^{\ast }\right) a}|_{\left( s_{2},t_{2}\right) =0}.  \tag{B.3}
\end{equation}

\textbf{(b) Normalization}

The normalization factor is
\end{subequations}
\begin{subequations}
\begin{equation}
N_{2}=e^{\left( g^{2}-1\right) \left\vert \alpha \right\vert ^{2}}\partial
_{s_{2}}^{m}\partial _{t_{2}}^{m}e^{gt_{2}\alpha +gs_{2}\alpha ^{\ast
}\allowbreak +\allowbreak s_{2}t_{2}}|_{\left( s_{2},t_{2}\right) =0},
\tag{B.4}
\end{equation}%
which leads to the analytical result of $N_{2}$ in Eq.(\ref{1-4}).

\textbf{(c) Expectation value}

Here, we give $\langle a^{\dagger k}a^{l}\rangle _{\rho _{2}}$\ as follows
\end{subequations}
\begin{subequations}
\begin{align}
\langle a^{\dagger k}a^{l}\rangle _{\rho _{2}}& =\frac{e^{\left(
g^{2}-1\right) \left\vert \alpha \right\vert ^{2}}}{N_{2}}\partial
_{s_{2}}^{m}\partial _{t_{2}}^{m}\partial _{f_{2}}^{k}\partial _{h_{2}}^{l}
\notag \\
& e^{g\left( h_{2}+t_{2}\right) \alpha +g\allowbreak \left(
f_{2}+s_{2}\right) \alpha ^{\ast }}  \notag \\
& e^{\allowbreak h_{2}s_{2}+\allowbreak f_{2}t_{2}+s_{2}t_{2}}|_{\left(
s_{2},t_{2},f_{2},h_{2}\right) =0}.  \tag{B.5}
\end{align}

\textbf{(d) Density matrix elements}

Here, we give $\rho _{kl}^{(2)}=\left\langle k|\rho _{2}|l\right\rangle $
with the following form
\end{subequations}
\begin{subequations}
\begin{align}
\rho _{kl}^{(2)}& =\frac{e^{-\left\vert \alpha \right\vert ^{2}}}{N_{2}\sqrt{%
k!l!}}\partial _{s_{2}}^{m}\partial _{t_{2}}^{m}\partial
_{f_{2}}^{k}\partial _{h_{2}}^{l}  \notag \\
& e^{f_{2}s_{2}+h_{2}t_{2}+g\left( f_{2}\alpha +h_{2}\alpha ^{\ast }\right)
}|_{\left( s_{2},t_{2},f_{2},h_{2}\right) =0}.  \tag{B.6}
\end{align}

\textbf{(e) Wigner function}

WF of $\rho _{2}$ has the following form
\end{subequations}
\begin{subequations}
\begin{align}
& W_{\rho _{2}}\left( \beta \right)  \notag \\
& =\dfrac{2}{\pi N_{2}}e^{-\left( g^{2}+1\right) |\alpha |^{2}-2\left\vert
\beta \right\vert ^{2}+2g\allowbreak \left( \alpha \beta ^{\ast }+\alpha
^{\ast }\beta \right) }  \notag \\
& \partial _{s_{2}}^{m}\partial _{t_{2}}^{m}e^{-g\left( t_{2}\alpha
+s_{2}\alpha ^{\ast }\right) -s_{2}t_{2}+\allowbreak 2\left( t_{2}\beta
+s_{2}\beta ^{\ast }\right) }|_{\left( s_{2},t_{2}\right) =0}  \tag{B.7}
\end{align}

\begin{acknowledgments}
This project was supported by the National Natural Science Foundation of
China (No. 11665013).
\end{acknowledgments}

\end{subequations}

\end{document}